\documentclass{iaus}
\usepackage{graphicx}

\title[Energetics of Coronal Mass Ejections] %% give here short title %%
{Energetics of Coronal Mass Ejections}

\author[Subramanian and Vourlidas]   %% give here short author list %%
{Prasad Subramanian$^1$,
 Angelos Vourlidas$^2$}

\affiliation{$^1$Inter-University Centre for Astronomy and Astrophysics, P.O Bag 4, Ganeshkhind, Pune - 411007, India \break email: psubrama@iucaa.ernet.in\\[\affilskip]
$^2$Code 7663, Naval Research Lab, Washington, DC 20375, USA \break email: vourlidas@nrl.navy.mil}

\pubyear{2004}
\volume{226}  %% insert here IAU Symposium No.
\pagerange{119--126}
\date{?? and in revised form ??}
\jname{Coronal and Stellar Mass Ejections}
\editors{A.C. Editor, B.D. Editor \& C.E. Editor, eds.}
\begin{document}

\maketitle

\begin{abstract}
We examine the energetics of the best
  examples of flux-rope CMEs observed by LASCO in 1996-2001. We find
  that 69\% of the CMEs in our sample experience a driving power in
  the LASCO field of view.  For these CMEs which are driven, we
  examine if they might be deriving most of their driving energy by
  coupling to the solar wind. We do not find conclusive evidence to
  support this hypothesis. We adopt two different methods to estimate
  the energy that can possibly be released by the internal magnetic fields of the CMEs. We
  find that the internal magnetic fields are a viable source of
  driving power for these CMEs.
\end{abstract}

\firstsection % if your document starts with a section,
\section{Introduction}

The energy budgets involved in Coronal mass ejection (CME)
propagation at heights $\gtrsim$ 2R$_{\odot}$ are indicative of the
manner in which the CME interacts with streamers and the solar wind
and also provide constraints on the energies required during the
initiation phase.
%\section{Our sample}
 We study the evolution of potential and kinetic energies of
39 individual FR CMEs between 1997 and 2001.  This comprises a
complete sample of the best examples of FR CMEs observed by LASCO in
1996-2001 (out of about 4000 events). We first generate mass images of each CME and then calculate the
potential and kinetic energies using a procedure similar to that described in Vourlidas et al. (2001). 
For 27/39 CMEs in our sample, the mechanical energy rises 
linearly with time, whereas 12/39 CMEs show no such 
trend.
For the 27/39 CMEs for which mechanical energy rises 
linearly with time, there is a clear external driving power.
%\begin{figure}
% \includegraphics[height=2.0in,width=4.0in]{f1.eps}
%  \caption{The mechanical
%  (i.e., kinetic + potential) energy for two representative CMEs
%  plotted as a function of time from initiation. The mechanical energy
%  for the CME on 2000/03/22 (left panel) increases linearly with time,
%  implying that there is a constant driving power on the CME as it
%  propagates outwards.}\label{}
%\end{figure}
\section{Source of driving power: solar wind, or internal magnetic energy?}
If CMEs are propelled
via coupling with the solar wind, larger CMEs 
should have larger driving powers, since they have larger areas
for interacting with the solar wind and would therefore be better
coupled with it.
For the CMEs which have a driving power, figure 1 shows a scatterplot of size versus driving power.
From figure 1, the correlation between the driving power and CME size is evidently poor, and we find 
little evidence to suggest that
larger CMEs have more driving power. 
This casts doubt on the hypothesis that these CMEs
 are powered by momentum coupling with the
ambient solar wind.
\begin{figure}
 \includegraphics[height=3.5in,width=4.0in]{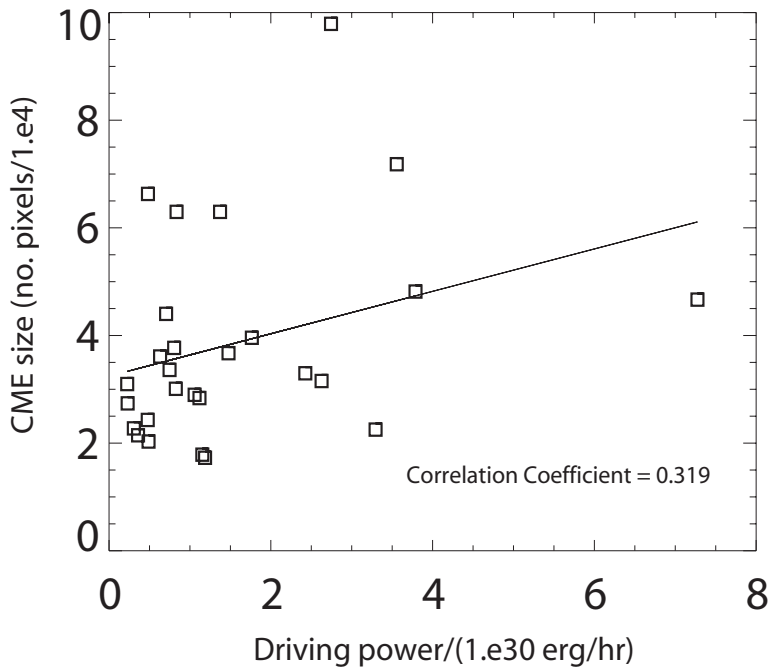}
  \caption{The mean size (in number of pixels) for CMEs 
  plotted as a function of their driving power. The low
  correlation coefficient suggests that there is no evidence to claim
  that larger CMEs have larger driving powers.}\label{}
\end{figure}

We have computed an estimate of the rate of energy released by the magnetic field advected by each CME using
two different methods. We envisage that the propelling force is
provided by some sort of $\vec{J} \times \vec{B}$ forces (due to misaligned magnetic fields and
currents) within the flux rope. One method uses the magnetic flux carried by near-earth magnetic 
clouds (e.g., Lepping et al. 1997) and assumes that this value is representative of the average
magnetic flux that is frozen into a typical CME in our dataset. According to this method, the internal
magnetic field of a CME can provide 0.744 $\pm$ 1.352 of the required driving power on the average. 
The other method uses 
direct estimates of the CME magnetic field from radio measurements (Bastian et al. 2001). According to
this method, the internal
magnetic field of a CME can provide $12.819 \pm 1.677$ of the required driving power on the average.
The details of these calculations can be found in Subramanian \& Vourlidas (2004).
\section{Conclusions}
We find no
evidence to suggest that the driven CMEs in our sample derive their driving power
primarily via coupling with the solar wind.
We employ two different approaches to investigate
if release of the internal magnetic energy in the CME can possibly provide the driving power.
One approach
suggests that, on the average, energy released by the internal magnetic field of a CME can provide upto
74\% of the required driving power. Another suggests that the energy released by the internal magnetic field of
a CME can be around an order of magnitude greater than what is required to drive it.

\end{document}